# An ultrafast polarised single photon source at 220 K

Tong Wang[1,†,*], Tim J. Puchtler[1,†,*], Tongtong Zhu[2], John C. Jarman[2], Luke P. Nuttall[1], Rachel A. Oliver[2], and Robert A. Taylor[1]

ABSTRACT: A crucial requirement for the realisation of efficient and scalable on-chip quantum communication is an ultrafast polarised single photon source operating beyond the Peltier cooling barrier of 200 K. While a few systems based on different materials and device structures have achieved single photon generation above this threshold, there has been no report of single quantum emitters with deterministic polarisation properties at the same high temperature conditions. Here, we report the first device to simultaneously achieve single photon emission with a $g^{(2)}(0)$ of only 0.21, a high polarisation degree of 0.80, a fixed polarisation axis determined by the underlying crystallography, and a GHz repetition rate with a radiative lifetime of 357 ps at 220 K. The temperature insensitivity of these properties, together with the simple planar growth method, and absence of complex device geometries, makes this system an excellent candidate for on-chip applications in integrated systems.

[1]Department of Physics, University of Oxford, Parks Road, Oxford, OX1 3PU, UK.
[2]Department of Materials Science and Metallurgy, University of Cambridge, 27 Charles Babbage Road, Cambridge, CB3 0FS, UK.
[†]These authors contributed equally to this work.
[*]email: tong.wang@physics.ox.ac.uk; tim.puchtler@physics.ox.ac.uk

Exploiting the quantum nature of light-matter interactions often relies on the presence of extreme conditions that cannot be applied outside the laboratory setting. In the case of polarised single photon sources[1], most of today's semiconductor systems require cryogenic temperatures below 20 K for the observation of sub-Poissonian photon statistics[2–9] and linearly polarised emission[10–15]. In order to move from proof-of-concept polarised single photon emitters towards integration into scalable electronic platforms, these non-classical semiconductor light sources need to break this temperature barrier and operate above the thermoelectric cooling threshold, typically around 200 K. Such devices would then be the key building-block for truly secure on-chip quantum communication[16–18] based on polarisation-reliant protocols, such as BB84[19], and act as qubit sources for linear optical quantum computing[20].

Although several systems have demonstrated unpolarised single photon emission above this temperature barrier[21–27], with some even reaching operation temperatures beyond ambient conditions[28], there have been no direct reports of single photon emission with deterministic polarisation properties above the 200 K threshold. Polarisation control alone has proven difficult to achieve, and has only been demonstrated at < 10 K via complicated device geometry manipulation[10] or the use of horizontally aligned nanowires[13], making electrical contacting laborious and integration of multiple devices or batch production inherently difficult. Similarly, high temperature single photon emission, regardless of polarisation control, is challenging to achieve. In order to produce an increased degree of quantum confinement to combat the thermal activation of non-radiative exciton decay pathways, it is often necessary to use both wide bandgap materials, such as III-nitrides[21–23,28], and structures with complex 3D growth routines, such as nanowire (NW)-quantum dots (QDs)[23,25,28], which again pose challenges for the development of electrically driven devices. Moreover, although the use of nitrides allows large band-offsets[29], the commonly used (0001) polar plane introduces the undesirable quantum-confined Stark effect (QCSE), thereby significantly reducing the exciton oscillator strength and decreasing the repetition rate and quantum efficiency. These complications and trade-offs in concurrently realising polarisation control, single photon emission, high temperature operation, and fast repetition rates, and the resulting complex device geometries, have significantly inhibited the incorporation of single photon emitters into realistically processable on-chip devices for applications in photonics.

In this work, we report the first device to simultaneously achieve on-demand single photon generation, a high degree of optical linear polarization, an identifiable predefined polarisation axis, and an ultrafast repetition rate at 220 K (−53 ºC). The strong light emission in the blue spectral region also enables use of the most efficient available detectors, at a temperature that can be reached easily by commercial Peltier coolers. Our sample consists of non-polar (11-20) *a*-plane InGaN QDs grown by metal-organic vapour phase epitaxy (MOVPE) embedded within a p-i-n doped GaN matrix, allowing future fabrication of electrically pumped devices. Nanopillar structures, such as that shown in Fig. 1a, were post processed for increased photon extraction efficiencies. The ability to grow InGaN/GaN QDs on the *a*-plane not only minimises the unwanted QCSE, increasing the exciton oscillator strength and radiative decay rate, enhancing quantum efficiency, and reducing coupling to phonon-assisted non-radiative exciton decay routes[30–32], but also causes highly polarised emission aligned to the [1-100] crystal axis[33]. Moreover, this high temperature ultrafast polarised single photon source requires only simple planar epitaxial growth, allowing devices to be fabricated easily using simple lithographic techniques.

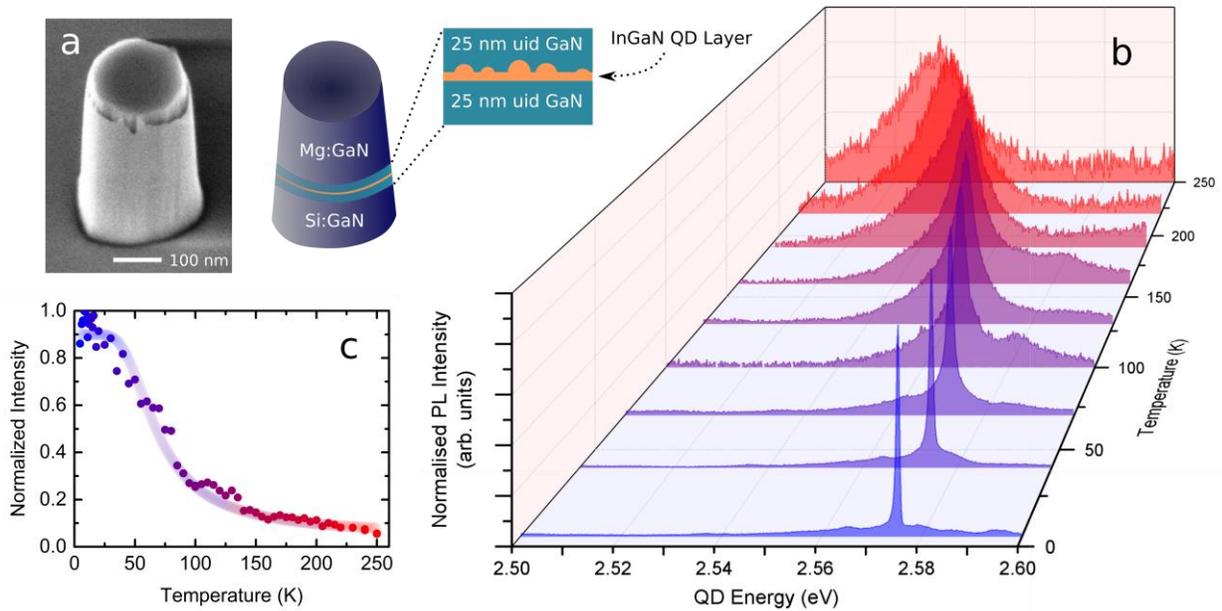

**Figure 1 | Temperature-resolved *µ*-PL of nanopillar-enhanced *a*-plane InGaN QDs. a,** Scanning electron microscope (SEM) image and schematic of a representative nanopillar structure in which QDs are embedded. **b,** Temperature dependent *µ*-PL spectra at 4.7 K, and from 40 to 250 K at 30 K intervals, with peak intensities normalized to the emission at 4.7 K. **c,** Integrated intensity from QD spectra between 4.7 and 250 K, at 5 K intervals. The plotted line shows an Arrhenius-type fit which well describes the decrease in emission intensity due to carrier escape[34].

The optical properties of the QDs were assessed by micro-photoluminescence (*µ*-PL) measurements under 800 nm excitation (see Methods). The wavelength of 800 nm was chosen to allow 2-photon excitation of the sample, as QDs have a larger relative absorption cross-section for multi-photon processes than quantum wells (QW), and hence the QD:background ratio is improved[35]. An example of the typical emission features can be seen in Fig. 1b, exhibiting a single sharp peak at ~ 2.58 eV with a full width at half-maximum (FWHM) of 722 ± 65 *µ*eV, attributed to the QD exciton and a low-intensity background caused by the fragmented QW underlying the QDs arising from the growth method[30]. As expected for a QD-bound exciton[36], the QD emission both redshifts and broadens in linewidth with increasing temperature, as evident in Fig. 1b, due to the well-known bandgap shrinkage and exciton-phonon coupling respectively. At 220 K, the emission energy has redshifted to 2.54 eV, and its FWHM increased to 19.0 ± 0.4 meV.

Furthermore, as thermally-assisted carrier escape becomes prominent at elevated temperatures, the emission intensity of the QD also decreases. As shown in Fig. 1c, the decrease in integrated intensity of the QD emission agrees very well with a standard single-channel semiconductor quenching model[34]. It is worth noting that the intensity decrease is slow, so that by 220 K, ~ 11% of the original intensity remains for the same excitation power. The strong QD emission at 220 K not only demonstrates operation well beyond the Peltier cooling barrier, but also allows us to study the QD's optical properties accurately.

In order to demonstrate the emission of single photons, autocorrelation experiments were performed using a Hanbury Brown and Twiss (HBT) method under pulsed excitation. For the studied QD, histograms of the variation in the number of coincidences with the delay time, τ, are proportional to the second order autocorrelation function

$$g^{(2)}(\tau) = \frac{\langle n(t)n(t+\tau)\rangle}{\langle n(t)\rangle\langle n(t+\tau)\rangle} \quad (1)$$

where $n(t)$ is the number of photons recorded by the detectors. Hence, the presence of an $n = 1$ Fock state for the verification of single photon emission requires the $g^{(2)}(0) < 0.5$ condition[37].

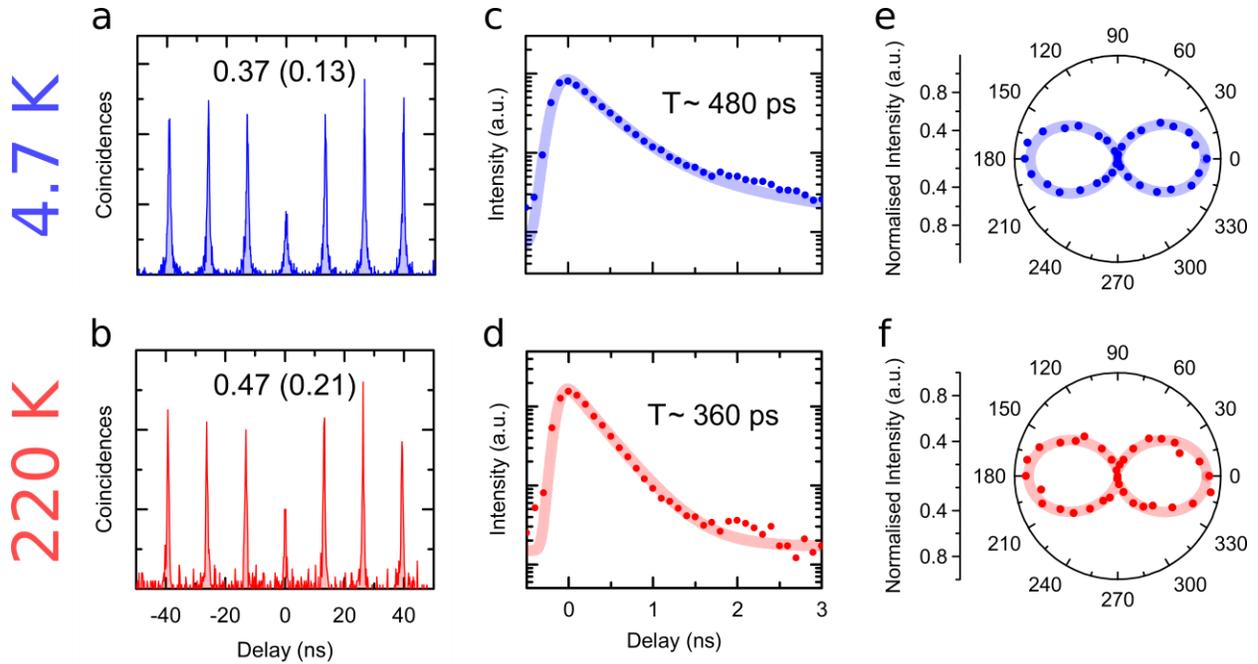

**Figure 2 | Experimental evidence of ultrafast polarised single photon emission at both 4.7 (blue) and 220 K (red). a,b,** Photon autocorrelation data, with raw and background-corrected $g^{(2)}(0)$ values demonstrating single photon emission. **c,d,** Time-resolved PL intensity plots of filtered QD signals from which the exciton decay constant is extracted. For greater accuracy, the fitted curves are double exponential functions convoluted with the near-Gaussian instrument response function. A slower, significantly weaker component becomes dominant after ~ 2 ns. **e,f,** Emission intensity variation with polariser angle. A Malus' law type sinusoidal fitting has been used to demonstrate that the emission is polarised, with the same polarisation axis at both low and high temperatures.

At 4.7 and 220 K, we measure raw $g^{(2)}(0)$ values of 0.37 and 0.47 respectively (Fig 2a and 2b), thus confirming the single photon nature of the emission. This is the first direct proof of single photon emission from an *a*-plane InGaN QD. Moreover, unlike other non-nitride single photon systems such as Ref. 24, the $g^{(2)}(0)$ values are very insensitive to temperature changes, and only increased slightly at 220 K. This could be attributed to the high exciton binding energies and large band offsets in III-nitride systems. For this reason, we have even observed ultrafast anti-bunching behaviour of this QD at 250 K (see Supplementary Information).

Non-zero raw $g^{(2)}(0)$ values are expected given the presence of spectrally overlapping background QW emission, as the QDs were formed on top of fragmented QWs during growth. Since single photon emission only comes from the QD; light of the selected wavelength from any other source present will increase the $g^{(2)}(0)$ values. To understand the $g^{(2)}(0)$ values expected from the QD alone, a commonly used background reduction calculation is performed. This background reduction can be accounted for using $\rho$, the ratio of QD intensity to the total intensity recorded by the PMTs, and the correction formula[38]

$$\frac{g^{(2)}_{\text{expt}}(0) - 1}{g^{(2)}_{corr}(0) - 1} = \rho^2. \tag{2}$$

The $\rho$ was measured to be 85% at 4.7 K, and 82% at 220 K, yielding $g^{(2)}_{corr}(0)$ values of 0.13 and 0.21 respectively (Fig. 2a and 2b). These values are much closer to 0, indicating that the QD is behaving as a pure SPS merely in the presence of a weak QW background. As we use non-resonant excitation, rapid repopulation and re-emission[39] from the QDs could be reasons for the occasional emission of more than one photon. The finite response time of the PMTs and the fast exciton radiative lifetime could also contribute to the non-zero $g^{(2)}_{corr}(0)$ values as explained below.

The characteristic radiative lifetime of the QD has been assessed by performing time-resolved μ-PL. The resulting decay plots at 4.7 and 220 K can be seen in Fig. 2c and 2d. Fitting of this lifetime data has been performed using a modified Gaussian function

$$f(t) = f_0 + (f_{\text{Gauss}} \otimes f_{\text{bi-exp}})(t) \tag{3}$$

where the instrument response function of the PMT detector (measured as a Gaussian with FWHM of 130 ps), $f_{\text{Gauss}}$, is convoluted with a bi-exponential decay $f_{\text{bi-exp}}$. The fitting gives a two-component exponential, with one intense fast component (~ 400 ps) attributed to the QD and another significantly less intense slow component (~ 4.0 ns).

The fast component has a decay constant of 480 ± 20 ps at 4.7 K, decreasing to 357 ± 20 ps at 220 K. Such a decrease in decay time with increasing temperature is seen in CdSe QDs[40], although arsenide QDs typically show uncertain temperature dependence of radiative lifetimes[41]. The closeness of these radiative lifetimes to the finite detector response time of 130 ps not only requires a modified Gaussian function for analysis accuracy, but also contributes to the non-zero $g^{(2)}_{\text{corr}}(0)$ values obtained in the previous section. There is a significant probability that the emission happened before 130 ps, during which time the PMTs could not distinguish the two events, thus adding to the histograms at $\tau = 0$. This is also at least part of the reason that the $g^{(2)}_{\text{corr}}(0)$ at 220 K is higher than that at 4.7 K, given the even faster radiative lifetime at higher temperatures.

These decay times are significantly faster than those reported in *c*-plane InGaN QDs, which are typically a few ns, supporting the assertion that the use of the *a*-plane orientation for these QDs reduces the internal field of the QDs, reducing the QCSE and allows higher oscillator strengths and faster repetition rates[30–32]. Furthermore, the lifetimes remain short and relatively temperature insensitive across the studied temperature range, promising reliable GHz repetition under Peltier cooled conditions.

The polarisation properties of the ultrafast single photon emission have been analysed by polarisation-resolved $\mu$-PL. Our previous investigations[33] have shown that the emission from an

*a*-plane InGaN QD not only has a statistically high average degree of optical linear polarisation (DOLP) of 0.90 ± 0.08, but also possesses an intrinsic axis of polarisation along the crystal *m*-direction. PL intensities of the studied QD at different polariser angles at both 4.7 and 220 K have been recorded and analysed. As shown in Fig. 2e and 2f, the sinusoidal fittings in accordance with Malus' Law show that the emission is polarised at both temperatures. The maximum and minimum intensities recorded, $I_{max}$ and $I_{min}$, were used to calculate the polarisation degree with the formula

$$\text{DOLP} = \frac{I_{max} - I_{min}}{I_{max} + I_{min}}. \qquad (4)$$

DOLP values of 0.83 ± 0.01 and 0.80 ± 0.13 have been obtained for 4.7 and 220 K respectively, indicating highly polarised single photon emission at both low and high temperatures. The use of *a*-plane QDs breaks the nitride wurtzite symmetry and lowers it to orthorhombic, leading to hole state splitting and band mixing effects[42]. The resultant exciton hole ground state has a much higher contribution from the state associated with emission along the *m*-direction than the *c*-direction, yielding not only high DOLPs, but also a deterministic polarisation axis[33,43]. Indeed, the axis of polarization for all QDs observed lies along the *m*-direction of the sample. As such, the direction of polarisation for these polarised single photon emitters is predefined by the material crystallography, a simpler approach than attempting to use strain engineering for polarisation control[10]. After sample preparation, striations are visible along the sample surface, arising from the growth process, and are aligned perpendicular to the *m*-direction, making the identification of the polarisation axis straightforward in our experiments. Furthermore, the DOLP at 220 K indicates that the QD-bound exciton transition has similarly high |*m*⟩-like

characteristics at elevated temperatures, thereby confirming the ability of *a*-plane InGaN QDs to operate as on-chip polarised single photon sources.

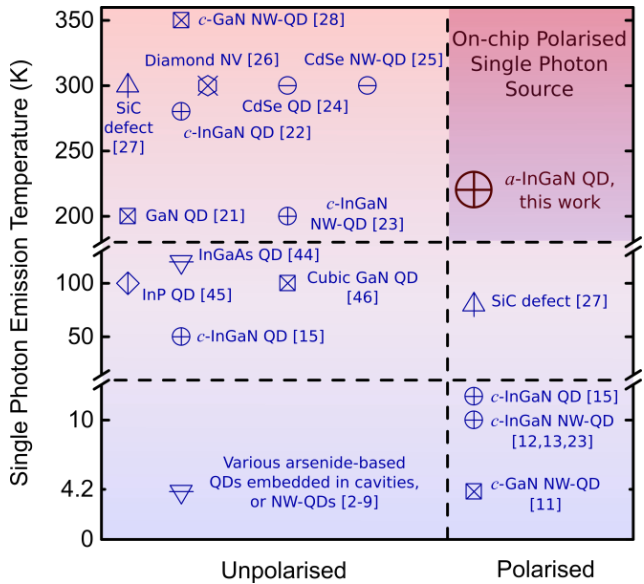

**Figure 3 | Comparison with other high temperature solid-state single photon sources.** Representative single photon devices fabricated with various materials and device structures. A few reports demonstrate high temperature single photon emission, but without directly showing polarised photon emission at such elevated temperatures. In these cases, such as Ref 13, two different markers have been used indicating single photon and polarised single photon emission separately, based on the results reported. For applications of on-chip polarised single photon sources, a device must be operating in the top right section of the figure.

Apart from the added advantages of a simple growth routine, planar structure, and ultrafast repetition rate, the two most important features for future device integration are simultaneous high temperature single photon emission and polarisation control. A review of state-of-the-art single photon emitters, to the best of our knowledge, is shown in Fig. 3, with the top right-hand section indicating the necessary condition for optimal on-chip single photon applications.

Although many single photon sources, such as the mature arsenide cavity-QD systems, have been able to produce highly indistinguishable near perfect single photon emission[2–9], their operations are limited to low temperatures, thereby precluding the possibility of thermoelectrically cooled on-chip operation on an electronic platform. Single photon emission at > 200 K is a challenge that only a few systems have achieved, as indicated in the top left-hand section of Fig. 3. However, most of the systems rely on NW-QD structures, SiC defect states, and diamond nitrogen vacancy centre engineering, which offer little prospect of electrically pumped devices. Furthermore, many of these devices have demonstrated either low temperature polarised single photon emission, or high temperature single photon emission, separately, which does *not* prove that they can be achieved simultaneously. The single photon emitters shown in the upper left-hand section, albeit reaching even beyond room temperature operation, lack the ability to generate polarised photons under the same conditions, unlike our *a*-plane InGaN/GaN QDs.

In summary, we have demonstrated simultaneous single photon generation and polarised light emission with a predefined polarisation axis, an ultrafast GHz repetition rate, and operation at 220 K, well above the thermoelectric cooling barrier for device integration into electronic systems. The radiative lifetime, polarisation degree and orientation, and $g^{(2)}(0)$ are all temperature insensitive. This breakthrough brings us closer to the realisation of on-chip polarised single photon sources.

**Method**

**Sample Preparation**

Non-polar (11-20) InGaN QDs were grown by a modified droplet epitaxy method as described previously[30]. InGaN QDs were positioned in the centre of a 50 nm thick intrinsic GaN layer, which was clad by 600 nm of n-doped GaN and 200 nm of p-doped GaN. Based on the analysis of atomic force microscopy (AFM) images of an uncapped InGaN QD sample, the QDs have an average height of ~ 7 nm, an average diameter of ~ 35 nm, and a density of approximately $1 \times 10^9$ cm$^{-2}$. To isolate individual QDs, we have processed the as-grown wafer into nanopillar structures by drop-casting of silica nanospheres onto the wafer as an etch mask, followed by dry etching to a depth of ~ 350 nm. The residual silica nanosphere etch mask was then removed by ultra-sonication and a buffered-oxide etch. Details of the p-i-n sample growth and nanopillars processing conditions can be found in Ref 33. Such nanopillar structures provide better collimation and directionality for the emitted photon, thus higher photon extraction efficiency.

**Optical Characterization:**

Micro-photoluminescence experiments were performed with the sample mounted on a nano-positioning system (~ nm precision Attocube positioners) contained in a closed-cycle cryostat (AttoDRY 800), varying the sample temperature between 4.7 and 300 K. Excitation was provided using a mode-locked Ti:Sapphire laser operating at a wavelength of 800 nm (pulse duration of ~ 1 ps, repetition rate ~ 76 MHz).

The excitation laser was focussed onto the sample through an objective lens (100×, 0.5 N.A.), with sample emission collected back through the same objective. Spectra are assessed using a 0.5 m focal length spectrometer (1200 l/mm grating). HBT experiments were performed by spectrally isolating the QD using a pair of tuneable bandpass filters and passing the filtered signal through a 50:50 beam-splitter connected to two PMTs. Signals from these PMTs are time-tagged using a time-correlated single-photon counting module with time resolution 25 ps. The PMTs define start and stop times individually. Similarly, lifetime data is collected by passing the spectrally filtered emission from the sample to a single PMT and using a fast-photomultiplier at the excitation laser as the start timer of the TCSPC module.

Polarization measurements were performed by introducing a linear polariser and half-wave plate into the optical collection arm of the $\mu$-PL system, with the transmission axis (0º marking) aligned to the PL component parallel to the [1-100] $m$-axis direction of the sample. The half-wave plate was used in order to maintain the same polarisation axis for light entering the spectrograph, in order to negate any polarisation dependent effects in the detection system itself.

# Supplementary Information:

# An ultrafast polarised single-photon source at 220 K

**Tong Wang[1,*], Tim J. Puchtler[1,*], Tongtong Zhu[2], John C. Jarman[2], Luke P. Nuttall[1], Rachel A. Oliver[2], and Robert A. Taylor[1]**

[1]Department of Physics, University of Oxford, Parks Road, Oxford, OX1 3PU, UK.

[2]Department of Materials Science and Metallurgy, University of Cambridge, 27 Charles Babbage Road, Cambridge, CB3 0FS, UK.

*tong.wang@physics.ox.ac.uk

*tim.puchtler@physics.ox.ac.uk

# QD PERFORMANCE AT 250 K

## PL Characteristics and QD:Background Estimation

The strong confinement and exciton binding energy of nitrides allow us to observe QD emission at temperatures higher than 220 K. Figure S1 displays the $\mu$-PL spectrum of the studied dot at 250 K, a temperature even closer to ambient conditions, and easier to reach for on-chip thermoelectric cooling. This is the highest operation temperature of $a$-plane InGaN QDs ever reported. Despite significant carrier recombination through non-radiative pathways, the sharp emission feature of the QD emerges from the QW background with a peak intensity of more than 50 counts/s on our $\mu$-PL system. Under the same excitation power, the integrated intensity of the QD remains ~ 6% of that at 4.7 K, which again shows the relatively slow thermal quenching of emission. At this temperature, the QD energy has redshifted to ~ 2.53 eV. Phonon-assisted homogeneous broadening has also increased the QD linewidth to $24.8 \pm 0.5$ meV. While a Voigt function that combines a Gaussian and a Lorentzian might provide a more accurate fitting, a single Lorentzian profile, as shown in Fig. S1, does indeed model the experimental data well enough. The underlying QW emission has been fitted to a Gaussian with a much wider linewidth of ~ 80 meV.

As an example of the background estimation method used in the main manuscript, the integrated intensity of the QD and the background QW, within the spectral window indicated in Fig. S1, have been calculated. The ratio of the QD intensity to the total intensity, $\rho$, can hence be estimated accordingly. In the 250 K case, the $\rho$ has been estimated to be ~ 57%. However, the difficulty in ascertaining the exact spectral coverage to be included in the intensity integration is a limitation of this estimation method. The band pass filters used in the spectral selection do not have perfect boxcar-like transmission profiles, adding uncertainty to the background estimation.

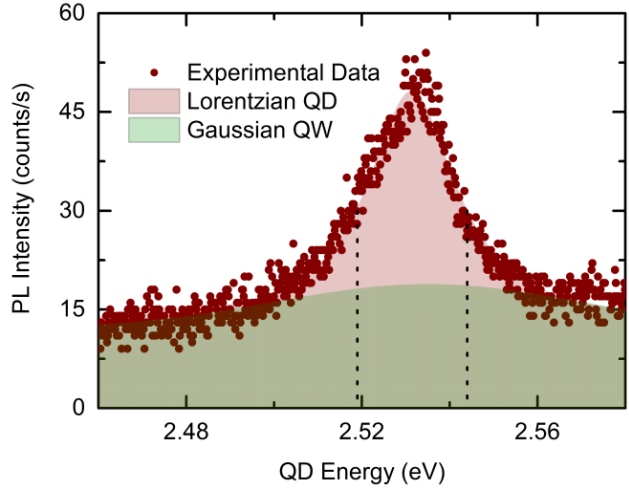

**Figure S1 | μ-PL of the studied QD at 250 K.** The sharp emission feature of the QD has been fitted to a Lorentzian, and the background QD emission to a Gaussian. The dotted line shows the spectral window with which the integrated intensities of the QD and QW emission have been estimated.

**Photon Antibunching**

The emission of the studied QD at 250 K has been passed to the HBT setup of the optical system, and the photon autocorrelation data recorded in Fig. S2. Despite the large amount of phonon-induced linewidth broadening and ~ 43% of background QW emission, the studied QD shows antibunching behaviour with a raw $g^{(2)}(0)$ value of 0.73. Although this uncorrected result is more than 0.5, and thus not sufficient to directly provide evidence for single photon generation, the $g^{(2)}(0)$ value can be corrected with equation (2) in the main manuscript text. With a $\rho$ of 57% as estimated earlier, the corrected $g^{(2)}(0)$ becomes 0.17, thereby indicating that the QD is indeed emitting as a single photon source at 250 K in the presence of underlying background QW emission. However, the uncorrected g[^(2)](0) should still be treated as the main performance indicator of the QD, and background reduction techniques would be needed for cleaner single photon emission at such elevated temperatures. 250 K is hence the highest temperature at which photon antibunching has been observed for *a*-plane InGaN/GaN QDs.

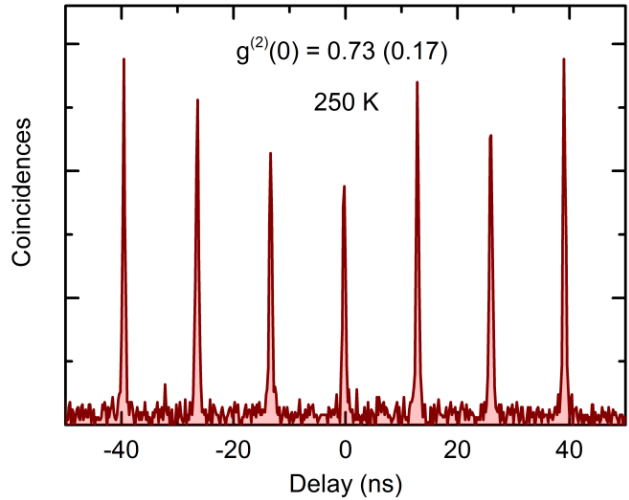

**Figure S2 | Autocorrelation data from HBT experiments of the studied QD at 250 K.** The value in the bracket is the background-corrected $g^{(2)}(0)$ value calculated using the method described in the main manuscript. The raw data provides direct evidence for antibunching behaviour, while the corrected $g^{(2)}(0)$ shows that the QD is emitting as a single photon source in the presence of overlapping QW emission.

**Radiative Lifetime**

Time-resolved $\mu$-PL has also been performed on the same QD to measure its radiative lifetime at 250 K. The fitting in Fig. S3 is performed using a modified Gaussian function, convoluted with an exponential decay, as described in the main manuscript. The decay constant has been found to be 191 ± 8 ps, indicating similarly strong electron and hole wavefunction overlap even at 250 K, and thus ultrashort radiative lifetime. The presence of a slower component after 0.6 ns, similar to that at 4.7 and 220 K, is still significantly weaker (< 10% intensity). Hence, the fast component is attributed to the QD emission, and agrees with expected radiative lifetimes of *a*-plane InGaN/GaN QDs. As such, we have shown strong, ultrafast, and antibunched photon generation from the studied QD at 250 K.

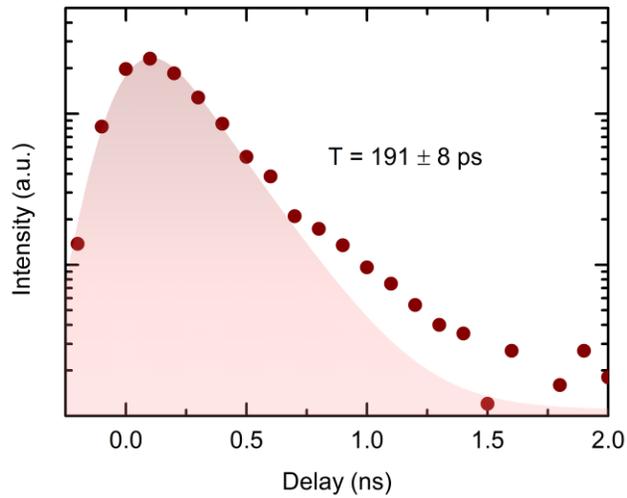

**Figure S3 | Radiative lifetime measurements with time-resolved *μ*-PL at 250 K.** TCSPC data has been fitted with a Gaussian function convoluted with an exponential. The lifetime calculation shows ultrafast antibunched photon emission at 250 K.